\documentstyle[11pt,moriond,epsfig]{article}

\def\Journal#1#2#3#4{{#1} {\bf #2}, #3 (#4)}

\def\PLB{{\em Phys. Lett.}  B}
\def\PRL{\em Phys. Rev. Lett.}
\def\PRD{{\em Phys. Rev.} D}
\def\ZPC{{\em Z. Phys.} C}

\def\be{\begin{equation}}
\def\ee{\end{equation}}
\def\bea{\begin{eqnarray}}
\def\eea{\end{eqnarray}}

\newcommand{\mett}{\mbox{${\rm \not\! E}_{\rm T}$}}
\newcommand{\xxbar}[1]{\mbox{$#1\overline{#1}$}}
\newcommand{\etal}{{ et al.}}

\newcommand{\twofig}[8]
{\begin{figure}[tb] \vspace{-1.5cm}
 \begin{minipage}[t]{0.48\linewidth}
   \epsfxsize \linewidth
   \epsffile[10 128 550 675]{#1}
   \vspace{#4}
   \vspace{1.0cm}
   \caption{#2}
   \label{#3}
 \end{minipage}
 \hfil
 \begin{minipage}[t]{0.480\linewidth}
   \epsfxsize \linewidth
   \epsffile[10 128 550 675]{#5}
   \vspace{#8}
   \vspace{1.0cm}
   \caption{#6}
   \label{#7}
 \end{minipage}
\vspace{-0.5cm}
\end{figure}
}

\newcommand{\twofigc}[8]
{\begin{figure}[tb] \vspace{0.5cm}
 \begin{minipage}[t]{0.48\linewidth}
   \epsfxsize \linewidth
   \epsffile[10 128 550 675]{#1}
   \vspace{#4}
   \vspace{1.0cm}
   \caption{#2}
   \label{#3}
 \end{minipage}
 \hfil
 \begin{minipage}[t]{0.480\linewidth}
   \epsfxsize \linewidth
   \epsffile[10 128 550 675]{#5}
   \vspace{#8}
   \vspace{1.0cm}
   \caption{#6}
   \label{#7}
 \end{minipage}
\vspace{-0.5cm}
\end{figure}
}

\begin{document}
\vspace*{4cm}
\title{NEW PHENOMENA II: RECENT RESULTS FROM THE FERMILAB TEVATRON}
\author{ DAVE TOBACK }

\address{Department of Physics, University of Maryland,\\
College Park, MD 20742, USA}

\maketitle\abstracts{The CDF and D\O\ collaborations continue to
search for new physics using more than 100~pb$^{-1}$ of \xxbar{p}
collisions at $\sqrt{s}=1.8$~TeV collected at the Fermilab Tevatron.
We present recent results from both experiments on R-parity violating
Supersymmetry and $Z'$/Technicolor production with $ee$ and
\xxbar{t}\ final states. In addition we introduce Sherlock, a new
quasi-model-independent search strategy.}

\section{Introduction}\label{Into}

\newcommand{\lickme}{
The Fermilab Tevatron collider run from 1992-1996 produced more than
120~pb$^{-1}$ of \xxbar{p} collisions at $\sqrt{s}=1.8$~TeV per
detector for both the CDF and D\O\ collaborations.  Since the
beginning of the run, the two collaborations have produced over 50
publications searching for new phenomena beyond the standard model.
While there has been no evidence for new physics, important limits
have been set on models of Supersymmetry, Higgs, Technicolor, extra
gauge bosons, compositness and others.

The results presented here include the first investigation of data
sets, the mining of published data sets for information on new
models, and new analysis techniques.

}

In these, and other\cite{Juan}, proceedings we summarize some recent
results of searches for new physics at the Fermilab Tevatron.
Specifically, we review new results on R-parity violating
Supersymmetry (SUSY) and $Z'$/Technicolor models in the $ee$ and
\xxbar{t}\ channels. We conclude by introducing Sherlock, a new
quasi-model-independent search strategy.

\section{R-parity Violating SUSY}

\twofig {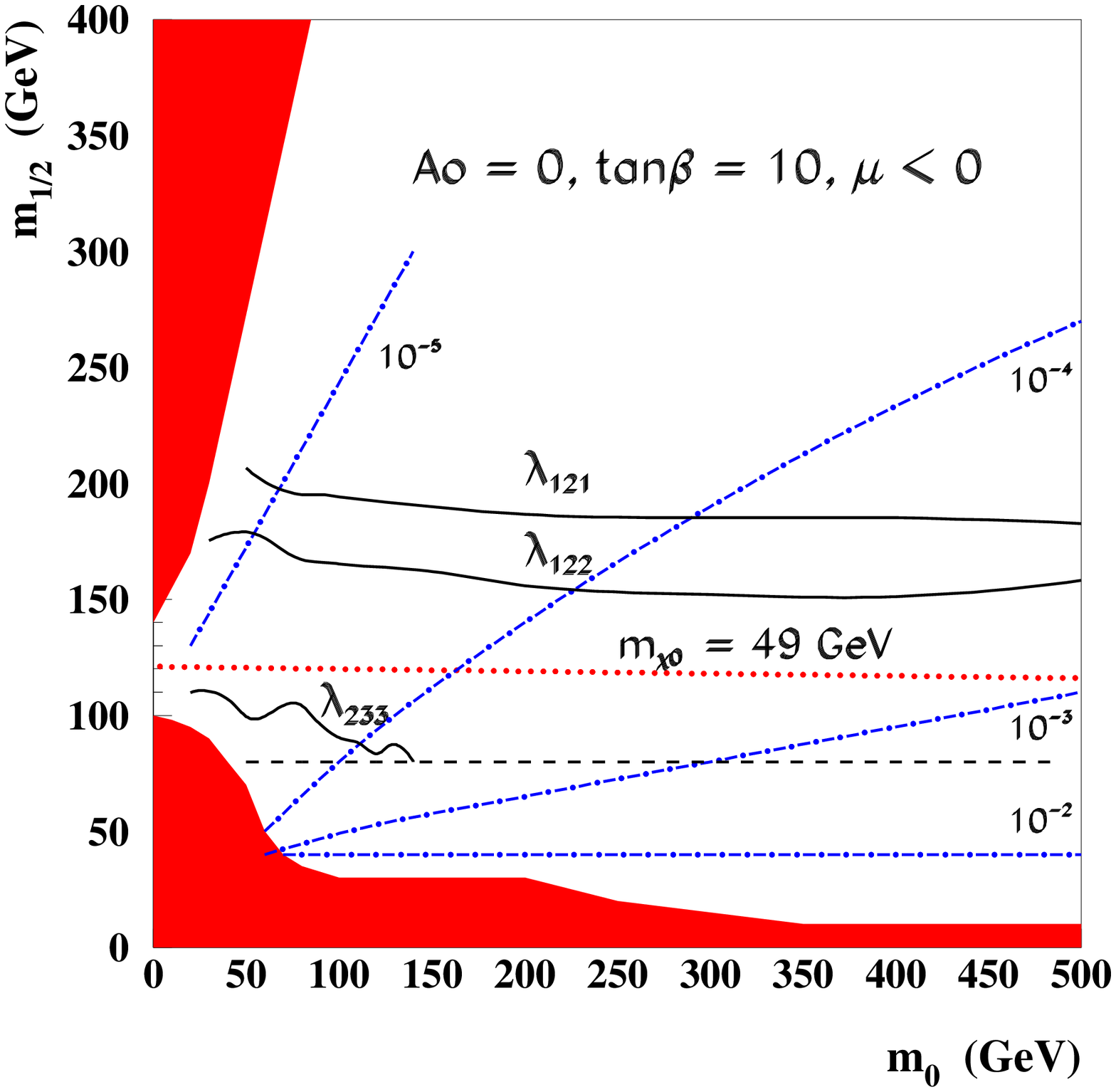}{Preliminary 95\% C.L. exclusion region on
all SUGRA R-parity violating SUSY production from D\O. The contours
are drawn for the value of the parameters $A_0=0$, tan$\beta$ = 10
and $\mu<$0, with finite values of $\lambda_{121}$, $\lambda_{122}$
and $\lambda_{233}$ couplings. For more detailed interpretation of
the lines see the text. }{Dzero RPV}{-0.0cm}
{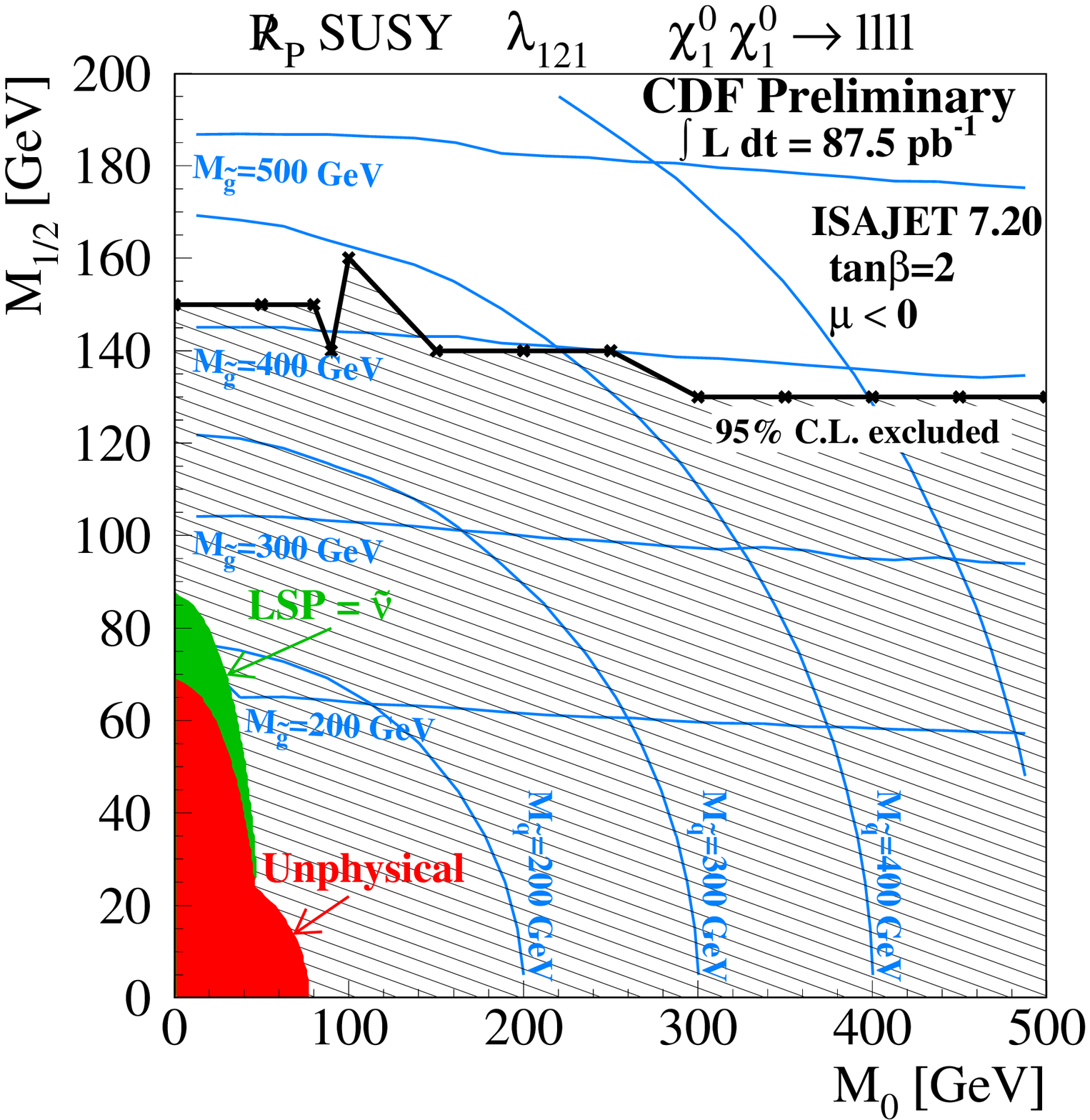}{Preliminary 95\% confidence level limit exclusion
region on R-parity violating SUSY from CDF. Here, the contours are
drawn for $A_0=0$, tan$\beta$ = 2, $\mu<$0, for the case of finite
$\lambda_{121}$ couplings. Note that not all SUGRA production is
considered in this analysis. }{CDF RPV}{-0.0cm}

Motivated in part by the interpretation\cite{HERA Theory} of the
reported HERA excess of high $Q^2$ events\cite{HERA Data}, and in
part for theoretical reasons, many recent searches for SUSY have
focused on the possibility that R-parity is not conserved. In such a
scenario at the Tevatron, pairs of SUSY particles will typically be
produced and then decay to standard model particles and two
neutralinos. However, instead of leaving the detector without
depositing any energy (as in R-parity conserving models), R-parity
violating terms, $\lambda_{ijk}\ell_i{\tilde \ell_j}{\bar e_k} +
\lambda '_{ijk}\ell_i{\tilde q_j}{\bar d_k} + \lambda_{ijk}'' {\bar
u_i}{\bar {\tilde d_j}}{\bar d_k}$, allow each neutralino to decay
via three body decay to standard model particles. For example in a
scenario with a non-zero all-leptonic coupling term, a neutralino
could decay via a $\nu_e$ and a virtual ${\tilde \nu_e}$ with the
${\tilde \nu_e}$ decaying via a $\lambda_{121}$ coupling and
producing $\mu^-e^+$. A similar decay of the other neutralino in the
event allows the event to produce a total of 4 charged leptons as
well as two neutrinos each of which can contribute to missing
transverse energy. The CDF and D\O\ collaborations have both searched
for this type of production and decay.

At D\O\ the search for R-parity violating SUSY is performed by
looking in the $eee, ee\mu, e\mu\mu$ and $\mu\mu\mu$ channels for
excesses of events with missing transverse energy. Since the search
is identical to that for ${\tilde \chi_1^{\pm}}{\tilde \chi_2^0}$
production and decay in mSUGRA, identical data sets and selection
criteria are employed\cite{Dzero WZ}. The results, along with the
luminosity for each sample, are shown in Table~\ref{Elemer Results}
with no candidates in the data. A similar search for R-parity
violating SUSY was performed at CDF in 87.5~pb$^{-1}$ of data. It is
complementary in that it requires four leptons, but does not require
missing transverse energy. Using all combinations of electrons and
muons in the final state, there is one candidate event in the data,
consistent with the background prediction of 1.3$\pm$0.4 events.

\begin{table}[htbp]
\centering
\begin{tabular}{|l|c|c|c|c|} \hline
Event categories &  $eee$  &  $ee\mu$  &  $e\mu\mu$  &  $\mu\mu\mu$
\\ \hline $L_{int}$  ($\mbox{pb}^{-1}$) &  $98.7\pm 5.2$  &  $98.7\pm
5.2$  &  $93.1\pm 4.9$  &  $78.3\pm 4.1$ \\ \hline Events observed &
0  &  0 &  0  &  0 \\ \hline Background   & $0.34\pm 0.07$  &
$0.61\pm 0.36$  &  $0.11\pm 0.04$  &  $0.20\pm 0.04$ \\ \hline
\end{tabular}
\caption{\small The result of the search for R-parity violating SUSY
in the trilepton signature at D\O. All errors are combinations of
both statistical and systematic errors.}

\label{Elemer Results}
\end{table}

Limits on R-parity violating models, which can be set as a function
of the mSUGRA SUSY model parameters ($m_0, m_{1/2}, A_0, tan\beta$
and $\mu$), are shown Figures~\ref{Dzero RPV} and \ref{CDF RPV} for
different values of the $\lambda$ coupling constant and $tan\beta$.
In both cases the regions of 95\% C.L. exclusion correspond to the
space below the dark solid lines. The lighter curves on
Figure~\ref{Dzero RPV} indicate the value of $\lambda$ such that the
average decay length of the LSP is less than 1 cm. Since the search
is not sensitive to displaced decays of the neutralino, the region
below the curves labeled with $\lambda_{121}$ and above the $10^{-3}$
line is excluded if $\lambda_{121}>10^{-3}$.

\newcommand{\lickmetwo}{In addition, in both plots there are two dark shaded areas, one
which indicates the regions where there is no electroweak symmetry
breaking, and one in which  the  neutralino is not the lightest SUSY
particle.}



\section{Technicolor and Neutral Heavy Vector Bosons}

\twofig {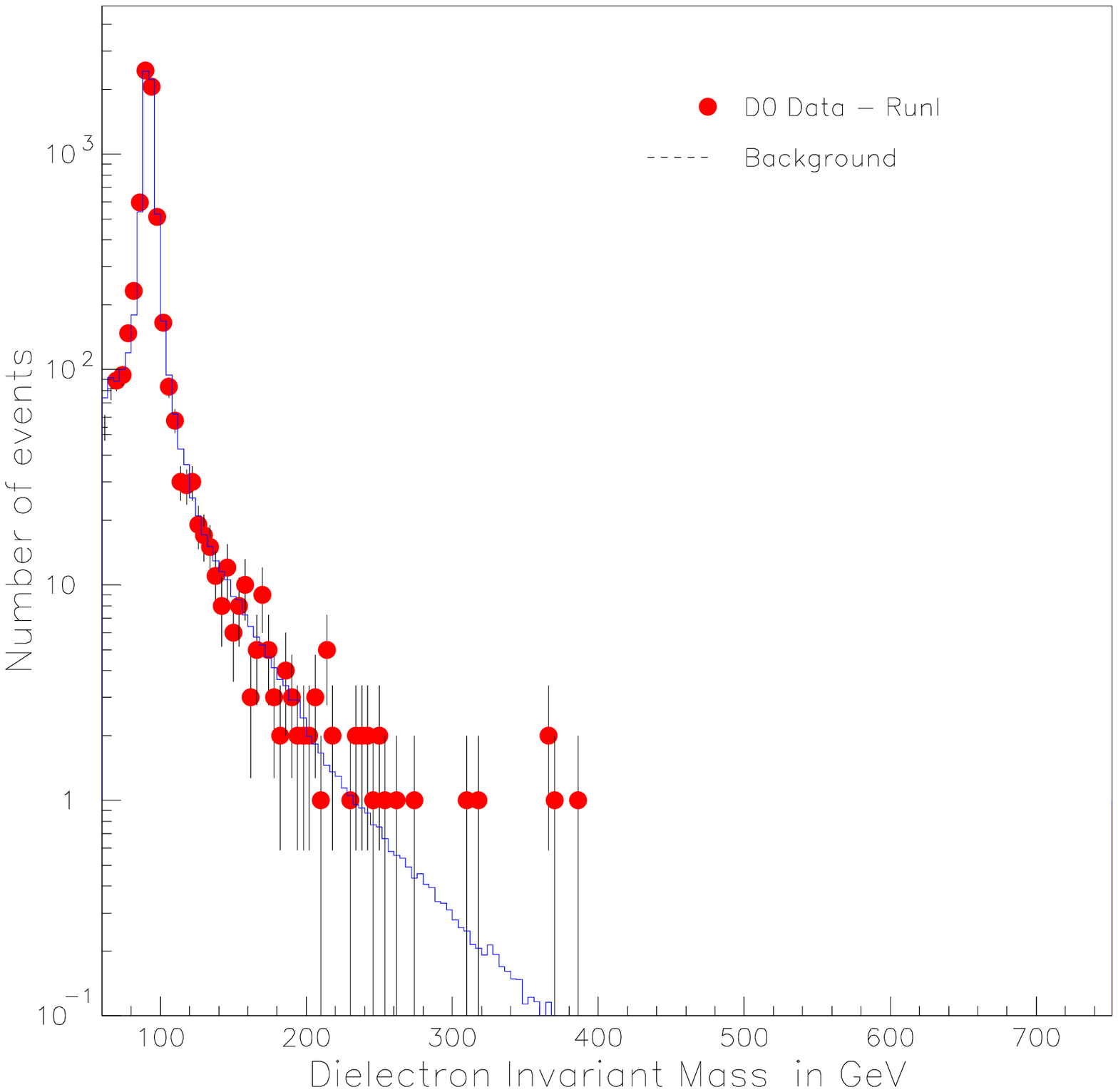}{The $ee$ invariant mass spectrum
from D\O. The background is normalized to the sum of the NNLO
$Z/\gamma^*$ production\protect\cite{Van Neerven} and fakes.}{ee data
plot}{0.0cm}
{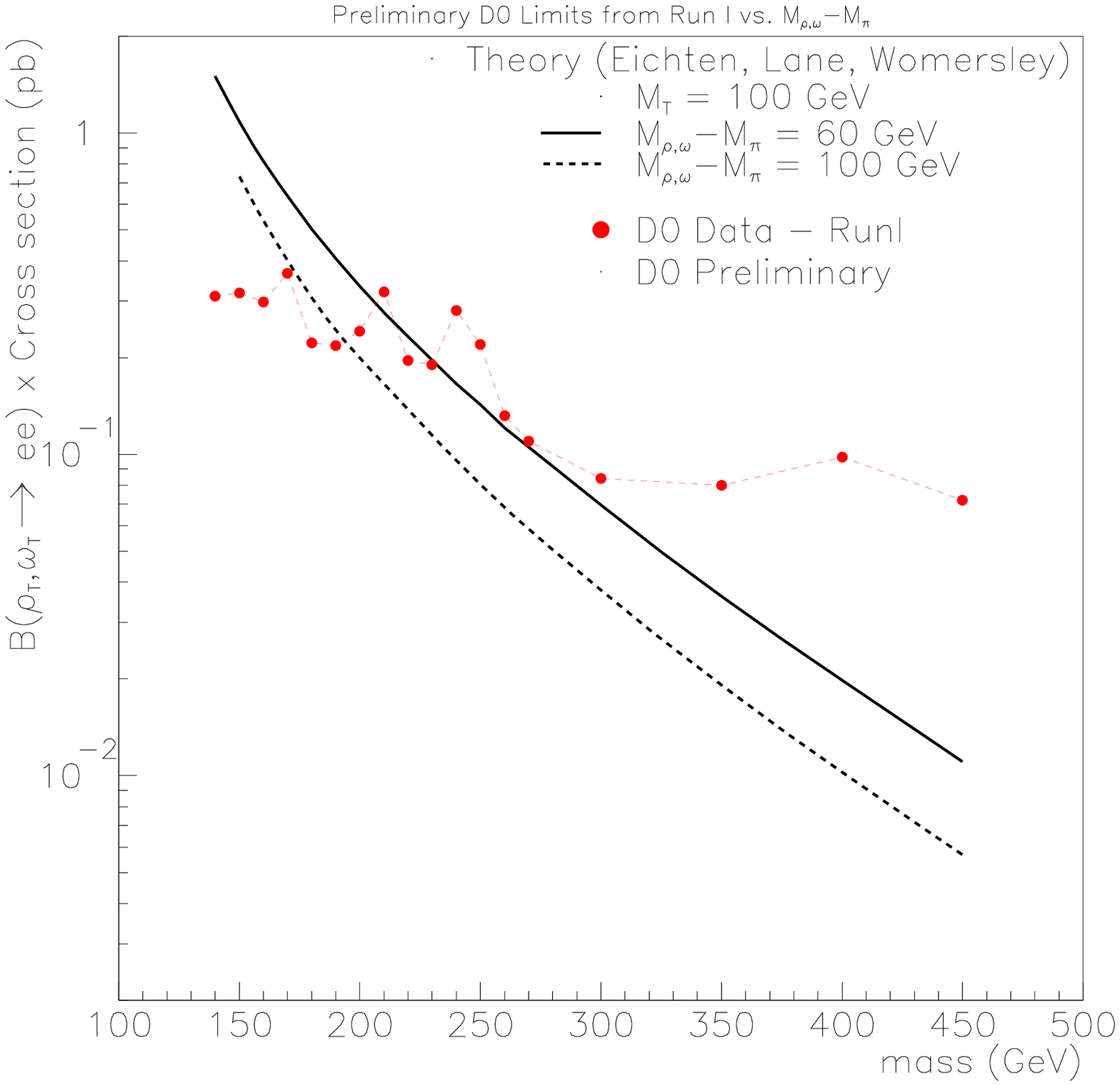}{D\O\ preliminary 95\% C.L. cross section
upper limits on production of $\rho_T/ \omega_T \rightarrow e^+e^-$.
The dots show the experimental cross section limit. The solid and
dashed lines show the theoretical cross section $\times$ branching
fraction for the combined $\rho_T$ and $\omega_T$ production.  The
two values correspond to whether the decay $\rho_T \rightarrow
W\pi_T$ is kinematically allowed.}{Techni Limit Plot}{0.0cm}

In Technicolor models\cite{TC Model} and models which have additional
neutral heavy vector bosons\cite{SM Zprime}, new heavy particles
(labeled $\rho_T, \omega_T$ or  $Z'$ respectively) are produced and
can decay via $e^+e^-$.  In both cases, a straight-forward search for
a resonance, or excess at high mass, in the $ee$ invariant mass
spectra could easily illuminate a signal. At D$\O$ this analysis uses
the same data set as the recently published paper setting limits on
quark and lepton compositeness\cite{Q&L Composite}. The luminosity
for this data set is 120.9~pb$^{-1}$. The backgrounds are dominated
by standard model production of $Z/\gamma^* \rightarrow ee$, and
instrumental backgrounds (fakes). The invariant mass spectrum for the
data and backgrounds is shown in Figure~\ref{ee data plot}. There is
no evidence for resonant production in the data or for an excess at
high mass. A similar search using both $ee$ and $\mu\mu$ final states
by CDF was recently published\cite{CDF Zprime Paper}. In both
searches there is no evidence for new physics.

Figure~\ref{Techni Limit Plot} shows the  D\O\ preliminary 95\% C.L.
cross section upper limits on production of $\rho_T/ \omega_T
\rightarrow e^+e^-$. Also shown on the plot are theoretical
production curves for two different scenarios in which the decay
$\rho_T \rightarrow W\pi_T$ is allowed or not allowed,  affecting the
branching ratio to $ee$. Assuming the $\rho_T$ and $\omega_T$ have
the same mass, the mass limits are M$_{\rho_T}>225$~GeV if the decay
$\rho_T \rightarrow W\pi_T$ is kinematically disallowed or if M$_{\rm
T} > 200$ GeV which suppresses the decay $\omega_T \rightarrow
\gamma\pi_T$. While CDF does not set specific limits on Technicolor
particles in this channel, their sensitivity is
comparable\cite{Kaori}.

Similarly, CDF and D\O\ also set limits on neutral heavy vector
bosons which decay via $Z'\rightarrow ee$. Assuming the couplings to
known fermions are the same as in the standard model, D\O\ sets a
mass limit of M$_{Z'}>670$~GeV in the electron only channel at 95\%
C.L. CDF's combined 95\% C.L. limit from $ee$ and $\mu\mu$  is
M$_{Z'}>$690~GeV. CDF goes further and sets limits on other $Z'$
models with extended gauge groups\cite{CDF Zprime Paper}.

Another search for neutral heavy vector bosons was carried out by the
CDF collaboration in the \xxbar{t} channel\cite{Juan}${}^, $\cite{CDF
tt Search}. The analysis, representing 106~pb$^{-1}$ of data, uses
the `lepton+jets' data set ($e$ and $\mu$) which was used for the top
quark mass measurements\cite{CDF Top}. Using the same fitting
techniques which are used to measure the mass, the final state
objects are fit to the \xxbar{t} hypothesis constraining the mass of
top quark mass to be 175~GeV. The best fit is then used to calculate
the invariant mass of the ${\bar t}t$ system which is searched for
resonant structure or excesses. There is no evidence for new physics
seen. While this channel is not competitive with $Z'$ limits from
$ee$ and $\mu\mu$ searches (assuming standard model couplings),
Technicolor models with leptophobic topcolor
production\cite{Topcolor} could make this channel highly produced at
the Tevatron. Experimental 95\% C.L. cross section upper limits are
at the few picobarn level and are compared with theoretical
production cross sections in Figure~\ref{CDF Zprime II}.

\twofigc {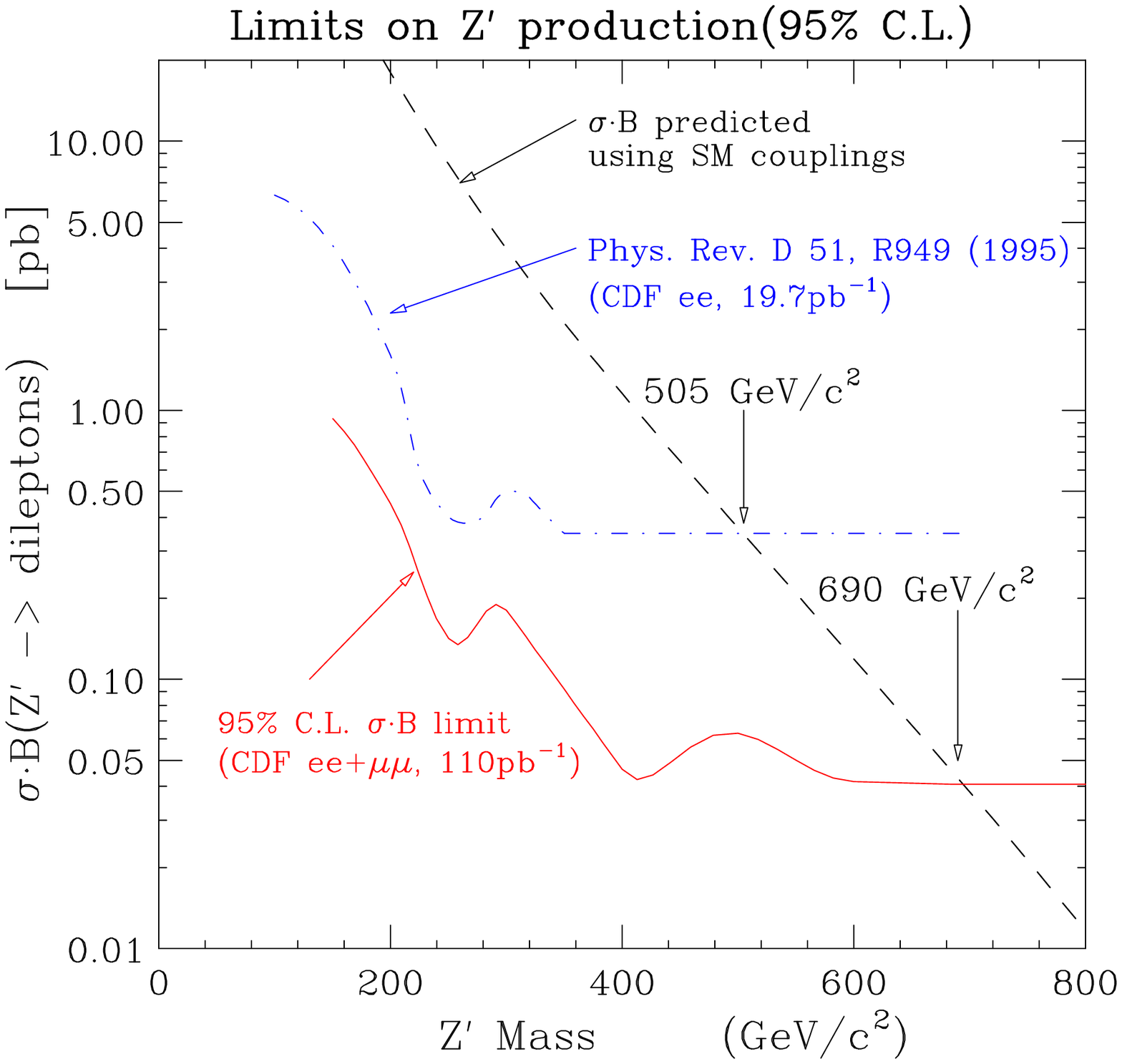}{CDF Published $Z'$ limit
plot\protect\cite{CDF Zprime Paper}. The solid line shows the 95\%
C.L. cross section upper limit as a function of the $Z'$ mass. The
dashed line shows the theoretical cross section $\times$ branching
fraction assuming standard model couplings for $Z' \rightarrow ee$.
}{CDF Zprime}{-1.5cm}
{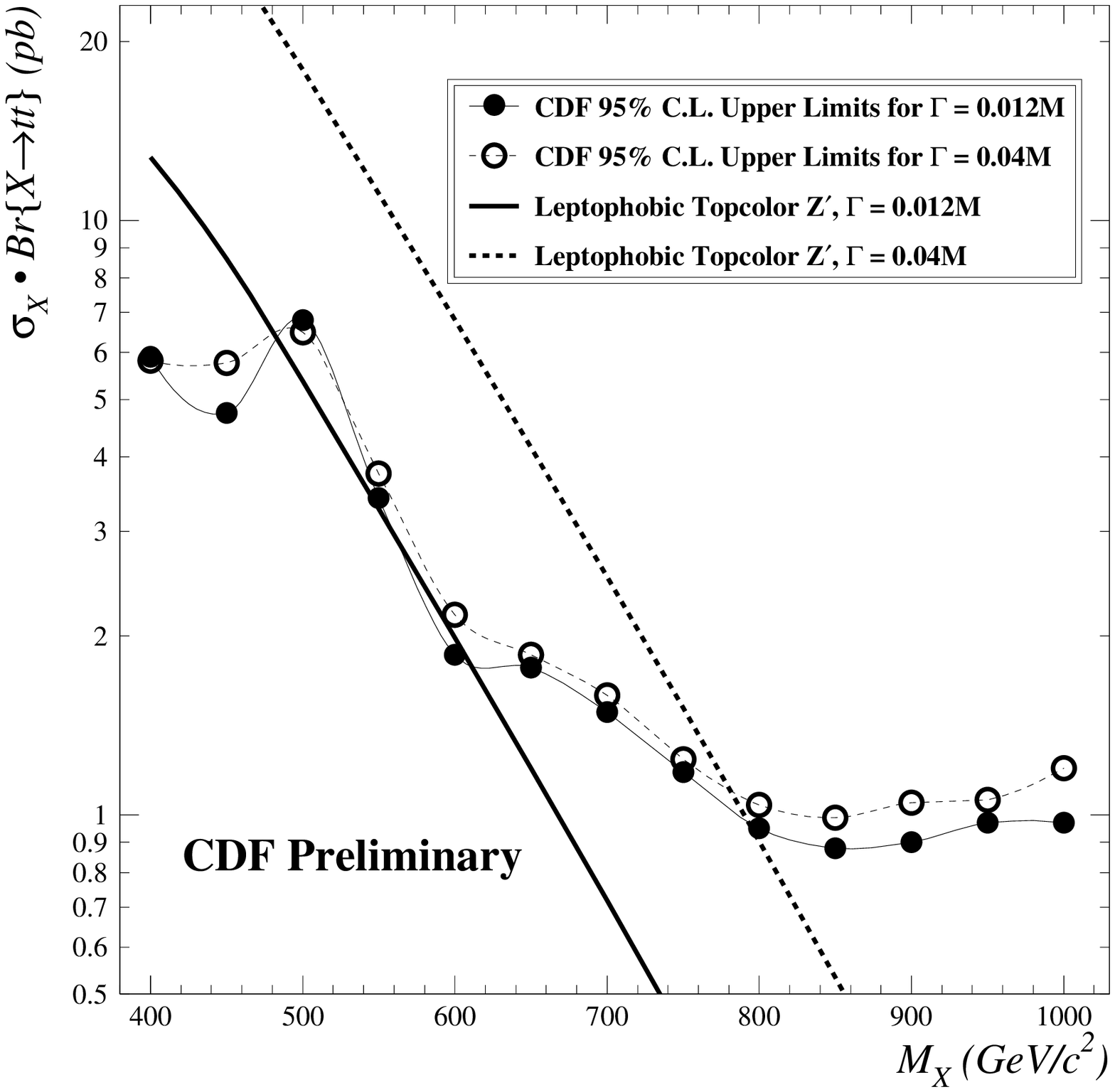}{CDF limits on $Z'\rightarrow$\xxbar{t} in
leptophobic topcolor theories\protect\cite{CDF tt Search}. The dots
show the 95\% C.L. cross section upper limit as a function of the
$Z'$. The solid and dotted lines show the theoretical cross section
$\times$ branching fraction for two values of the total width
$\Gamma$ of the $Z'$ in leptophobic topcolor models.}{CDF Zprime
II}{-1.5cm}

\section{Sherlock: A New Quasi-Model-Independent Method for Searches for New \mbox{Physics}}

\twofig {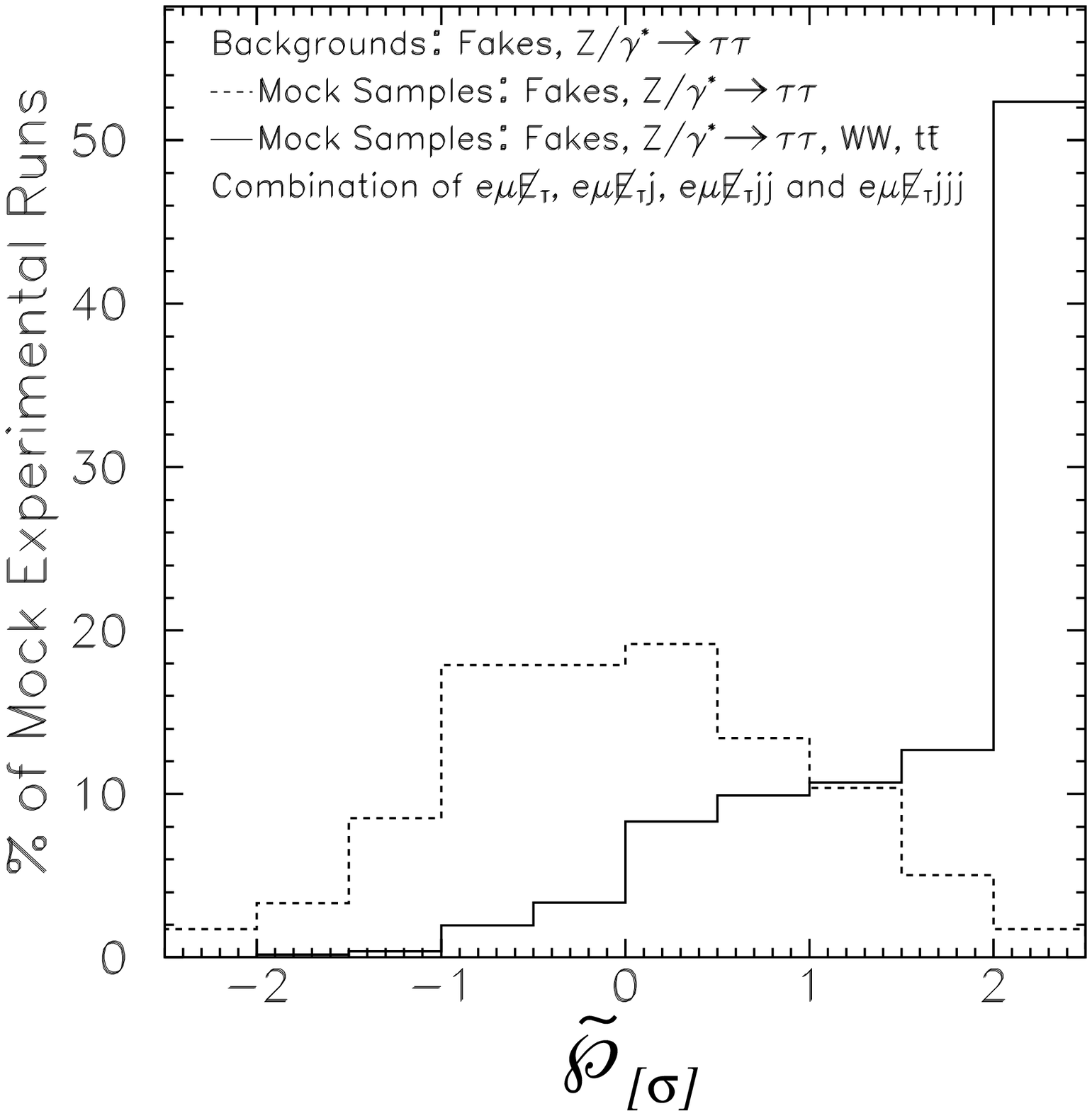}{The significance of the region of
greatest excess in standard deviations, denoted
$\tilde{P}_{[\sigma]}$, using Sherlock for an ensemble of mock data
experiments using $e\mu X$ for two cases: with a signal (solid line)
and with no signal (dashed line). In both cases the backgrounds are
$Z/\gamma \rightarrow \tau\tau$ and fakes. In the `no signal' case
the `data' is drawn only from the background distributions. As
expected, the results are basically Gaussian and centered on
0$\sigma$ with unit width. When `signal' events of WW and \xxbar{t}\
are added to the `data samples' but not to the background, Sherlock
reports that in the over 50\% of the mock experiments it finds a
statistically significant excess at or greater than 2$\sigma$. All
samples with $\tilde{P}_{[\sigma]}
>2\sigma$ are in the right most bin.}{Sherlock 1}{0.0cm}
{exer_wwttbar_twidpsig.eps}{The significance of the region of
greatest excess in standard deviations, denoted
$\tilde{P}_{[\sigma]}$, using Sherlock for an ensemble of mock data
experiments using $e\mu X$ as in Figure~\protect\ref{Sherlock 1}.
Here the backgrounds are $Z/\gamma \rightarrow \tau\tau$, fakes and
WW.  When `signal' events of \xxbar{t}\ are added to the `data' but
not to the background, Sherlock reports that in greater than 25\% of
the mock experiments it finds a statistically significant excess at
or greater than 2$\sigma$. All samples with $\tilde{P}_{[\sigma]}
>2\sigma$ are in the right most bin.}{Sherlock 2}{0.0cm}

Finally, we introduce Sherlock, a new quasi-model-independent search
method developed at D\O. This method provides a prescription for
searching for new physics by systematically looking for excesses in
multi-dimensional data distributions. Since we assume that the
physics responsible for electroweak symmetry breaking occurs at mass
scales large compared to standard model backgrounds, we currently add
in the assumption that the new physics is characterized by high
P$_{\rm T}$ final state particles. It is this feature which makes it
quasi-model-independent.

The method consists of a three part prescription and algorithm. The
first part is to pick a data set (such as the inclusive $e\mu$+X
sample) and categorize each event according to its observed final
state particles (number of electrons, jets, photons etc.). For each
category of events, the kinematic variables for the sample are
uniquely specified by an {\it a priori} prescription. A region, R, is
then defined in the multi-variable space surrounding one or more of
the data points.  By giving a precise definition of the region for an
event, or set of events, an amount of parameter space is determined
and the probability for the background in that region to fluctuate up
to or above the number of observed events in the region gives a
quantitative measure of the degree of interest of the region. The
algorithm then searches for the most interesting region (largest
excess relative to background) in all of variable space including the
high P$_{\rm T}$ region. Once this most interesting region is found,
it is compared to the most interesting region found in a large sample
of hypothetical similar experiments (HSEs) drawn from the parent
distributions of the backgrounds (according to statistical and
systematic uncertainties). In this way, the true degree of interest
of the region of largest excess is quantified in terms of the
fraction of HSEs which give regions which are more interesting than
the one observed in our data (again due simply to statistical
fluctuations or systematic misunderstanding of the data).

Sherlock is run on 108~pb$^{-1}$ of inclusive high P$_{\rm T}$
$e\mu$+X data taken at D\O. As a test and an illustration of the
sensitivity of the method, we have used this algorithm on a set of
mock experiments drawn from the background estimations. The $e\mu X$
sample has the advantage of having two known `signals' which give
high P$_{\rm T}$ physics in the final state: WW and \xxbar{t}\
production. Figure~\ref{Sherlock 1} shows the results of running a
series of mock experiments in which the `known' backgrounds include
only $Z/\gamma \rightarrow \tau\tau$ and fakes and the `data' is
drawn only from the background distributions. The results are shown
in the figure with the dashed lines which shows the significance in
standard deviations. As expected, the results are basically Gaussian
and centered on 0 with unit width. When `signal' events of WW and
\xxbar{t} are added to the `data' (according to expected standard
model production expectations, smeared by statistical and systematic
uncertainties) but not to the background expectations, as shown by
the solid line in Figure~\ref{Sherlock 1}, Sherlock reports that in
the over 50\% of the mock experiments it finds a statistically
significant excess at or greater than 2$\sigma$\footnote{We again
note that Sherlock doesn't know anything about $WW$ or \xxbar{t} and
that it is in no way optimized for finding them. It is simply looking
for an excess of events in the high-P$_{\rm T}$ region.}. Running
Sherlock on the data itself, using on $Z/\gamma \rightarrow \tau\tau$
and fakes as the background, Sherlock picks out significant excesses
($>2\sigma$) in both the $e\mu\mett$ ($WW$) and $e\mu\mett jj$
(\xxbar{t}) data correctly indicating the presence of both WW and
\xxbar{t} in the data.

Figure~\ref{Sherlock 2} shows the results when WW are added to the
background expectations so that we may see the sensitivity to
\xxbar{t}\ alone. Again, in an ensemble of mock experiments Sherlock
picks out an excess in the data at greater than or equal to the
2$\sigma$ level in over 25\% of the cases. Running on the data we
observe an excess at the 1.9$\sigma$ level in the $e\mu\mett jj$
data, correctly identifying \xxbar{t}\ in the data.

Including all known standard model sources in the backgrounds and
running Sherlock yields no evidence of an excess in any $e\mu X$
channel indicating agreement with the standard model. Specifically,
in 71\% of hypothetical similar experiments we expect to see an
excess more interesting than the most interesting region of excess
than is observed in our data.

The Sherlock method is a novel approach to searching for new physics
in the data. While a dedicated search is always best for a specific
signal hypothesis (e.g. SM Higgs with M$_H$=130~GeV), the number of
possible models to search for is very large. Sherlock is a very
powerful method for being sensitive to a large number of new physics
models by being much more model independent. While we have only
applied the method to a single data set, it is generally applicable
and should prove to be an immensely valuable tool in Run II to
complement the dedicated searches.

\section{Conclusions}

We have presented the results of a number of recent searches for new
physics from the Fermilab Tevatron based on the data taken during
1992-1996. While the next run with upgraded detectors and high
luminosity is just over a year away, we continue to make progress in
searching for new physics as well as setting limits on important
theoretical models. The future at the Tevatron appears bright, and
the lab should continue to be an interesting and exciting place to
search for new physics in the coming years.

\section*{Acknowledgments}

The author would like to thank Bruce Knuteson, Juan Valls, Maxwell
Chertok, Meenakshi Narain, Elemer Nagy, Gustaaf Brooimans, Andre
Turcot, Greg Graham, Ray Culbertson and Kaori Maeshima for their
assistance with this talk and proceedings.

\end{document}